\documentclass[twocolumn,twoside,slac_two]{revtex4}
\usepackage{graphicx}
\usepackage{fancyhdr}
\begin{document}

\title{Light Hidden Fermionic Dark Matter in Neutrino Experiments}

%

\author{Jennifer Kile}
\affiliation{Brookhaven National Laboratory, Upton, NY 11973, USA}

\begin{abstract}
We consider, in a model-independent framework, the potential for observing dark matter in neutrino detectors through the interaction $\bar{f} p \rightarrow e^+ n$, where $f$ is a dark fermion.  Operators of dimension six or less are considered, and constraints are placed on their coefficients using the dark matter lifetime and its decays to states which include $\gamma$ rays or $e^+e^-$ pairs.  After these constraints are applied, there remains one operator which can possibly contribute to $\bar{f} p \rightarrow e^+ n$ in neutrino detectors at an observable level.  We then consider the results from the Super-Kamiokande relic supernova neutrino search and find that Super-K can probe the new physics scale of this interaction up to $O(100\mbox{ TeV})$.
\end{abstract}

\maketitle

\thispagestyle{fancy}


\section{Introduction}

So far, many details about the nature of dark matter (DM) have eluded observational confirmation.  We do not yet know the number of DM species, their masses, or how they interact with each other or with Standard Model (SM) particles.  Additionally, there has been much recent interest in models with ``hidden'' sectors \cite{Strassler:2006im,Patt:2006fw,Han:2007ae,ArkaniHamed:2008qn}, where there exist light particles beyond the Standard Model (BSM) which have not yet been discovered because they are coupled to the SM only through interactions at some very high energy scale.  We can apply this idea to DM and consider the possibility that some or all of DM could be light (i.e. below the weak scale).  Thus, it makes sense to consider any easily observable possible interaction between light DM and SM particles and, if possible, in a model-independent way.  Here, we investigate a particular interaction which could be relevant for DM direct detection.  This work is based largely on \cite{Kile:2009nn}.

In the usual DM direct-detection scenario, a relatively heavy $O(10\mbox{ GeV}-10\mbox{ TeV})$ DM particle scatters elastically off of an SM particle, such as a nucleus.  If we take, as an example, a $100$ GeV DM particle scattering off of a $100$ GeV nucleus, the momentum deposited in the detector is on the order of the initial momentum of the DM particle, $O(100 \mbox{ MeV})$.

Instead, here we consider an interaction of the form 
\begin{equation}
f N \rightarrow f'  N'
\label{eq:proc}
\end{equation}  
where $f$ is a DM particle, $N$ and $N'$ are SM particles (here, nucleons), and $f'$ is some other particle which could be either contained within the SM or from BSM physics.  We consider the case where the $f'$ mass is much less than that of $f$, $m_f>>m_{f'}$.  In this case, the momenta of the final-state products will be of order $m_f$; thus, it seems plausible that existing experiments might be able to detect light DM if it scatters very inelastically.

Although one may consider the cases where $f'$ is invisible (i.e., a neutrino or a BSM particle), we take $f'$ to be a visible SM particle, an electron.  This interaction then looks much like an SM charged-current neutrino interaction; it is thus conceivable that neutrino experiments which can detect neutrinos of a given energy $E_{\nu}$ could be used to search for the $f$ particle, if $m_f\sim E_{\nu}$.  Although we initially only assume $m_f$ to be below the weak scale, when we insist that $f$ be appropriately long-lived, we find that the least constrained mass range is $m_f\lesssim O(100\mbox{ MeV})$, which can be probed by solar and reactor neutrino experiments.

For this analysis, we specifically examine the relevance of the Super-Kamiokande relic supernova neutrino search \cite{Malek:2002ns} to DM detection.   As we wish to be model-independent, we study the process $\bar{f} p \rightarrow e^+ n$ via an effective operator analysis.  We impose limits on these operators by insisting that the DM particle $f$ be long-lived  and rarely decay to easily-observable final states containing $\gamma$ rays or $e^+e^-$.  Finally, we discuss the cross-section for $\bar{f} p \rightarrow e^+ n$ in Super-K.  Super-K places a limit on the flux of $\bar{\nu}_e$'s via the process $\bar{\nu}_e p \rightarrow e^+ n$; we then use this result to place constraints on the process $\bar{f} p \rightarrow e^+ n$ and find that the new physics reach of Super-K to probe this process is $O(100\mbox{ TeV})$.  

The rest of this paper is organized as follows.  In Section \ref{sec:ops}, we give the criteria which our operators must satisfy and list our operator basis.  The DM lifetime and decays are used to place limits on these operators in Section \ref{sec:lim}.  Finding one operator which is not so tightly constrained, we give the experimental signatures and cross-section for $\bar{f} p\rightarrow e^+ n$ via this operator, and then put constraints on it using results from Super-K in Section \ref{sec:nus}.  We briefly discuss our results in Section \ref{sec:dis}, and, in Section \ref{sec:con}, we conclude.

\section{Operator Basis}
\label{sec:ops}
In this section, we enumerate the operators which we use for this analysis.  Here, $f$ is taken to be a fermion and a singlet under the SM gauge group $SU(3)\times SU(2)\times U(1)$.  We consider all operators of dimension six or less which are invariant under the SM gauge group and contribute to $\bar{f} p\rightarrow e^+ n$.  Redundant operators are then eliminated using integration by parts and the equations of motion for the fields.  Operators which contribute at tree level to neutrino mass would be very strongly constrained and are thus not included in the analysis.  We are thus left with six operators  
\begin{eqnarray}
{\cal O}_{W} &=& g \bar{L} \tau^a \tilde{\phi} \sigma^{\mu\nu} f W^a_{\mu\nu}\nonumber\\
{\cal O}_{\tilde{V}} &=& \bar{\ell}_R\gamma_{\mu}f \phi^{\dagger} D_{\mu} \tilde{\phi}\nonumber\\
{\cal O}_{VR} &=& \bar{\ell}_R \gamma_{\mu} f \bar{u}_R \gamma^{\mu} d_R, \\
{\cal O}_{Sd} &=& \epsilon_{ij }\bar{L}^i  f \bar{Q}^j d_R\nonumber\\
{\cal O}_{Su} &=& \bar{L} f \bar{u}_R Q \nonumber\\
{\cal O}_{T} &=& \epsilon_{ij }\bar{L}^i \sigma^{\mu\nu} f \bar{Q}^j \sigma_{\mu\nu} d_R \nonumber
\end{eqnarray}
where $L$ and $Q$ are the left-handed lepton and quark $SU(2)$ doublets of the SM, $\ell_R$, $u_R$, and $d_R$ are the right-handed singlets, $\phi$ is the SM Higgs field, and $\tilde{\phi}= i \tau^2 \phi^*$.  All of these operators are dimension-six and thus suppressed by $\Lambda^2$, where $\Lambda$ is taken to be some new physics scale above the weak scale.  Each of the operators ${\cal O}_I$ will also be accompanied by a coefficient $C_I$.  We note that, in all of the operators ${\cal O}_I$, $f$ is right-handed.

\section{Limits from Dark Matter Lifetime}
\label{sec:lim}
In order for the $f$ particle to be DM, it must satisfy at least a few basic constraints.  First of all, for DM to currently make up a significant fraction of the energy-density of the universe, it must decay to SM particles on a timescale at least on the order of the age of the universe, $4\times 10^{17}$ s.  Also, in order for DM to be sufficiently ``dark'', it must annihilate or decay to states containing easily-visible SM particles sufficiently slowly to have not yet been observed.  As our operators can contribute to decays of the $f$, we apply these decay constraints, considering each of our operators in turn.  

\begin{subsection}{${\cal O}_{W}$}
The operator ${\cal O}_{W}$ gives the decay $f\rightarrow \nu \gamma$ at tree level.  The width for this process is
\begin{equation}
\Gamma(f\rightarrow\nu\gamma) = \frac{|C_W|^2}{\Lambda^4} \frac{\alpha v^2}{2} m_f^3
\end{equation}
where $v$ is the SM Higgs vacuum expectation value (vev) and $\alpha$ is the SM fine-structure constant.  The authors of \cite{Yuksel:2007dr} find that DM which decays to two daughters, one of which is a photon, must have a lifetime $\gtrsim 10^{26}$ s for DM masses between $\sim 1$ MeV and $\sim 100$ GeV.  Taking $m_f = 1$ MeV (approximately the minimum value of $m_f$ which could be observable at neutrino experiments), we obtain  
\begin{equation}
\label{eq:ownum1}
\frac{|C_W|^2}{\Lambda^4}\lesssim \frac{1}{(8\times 10^7 \mbox{TeV})^4}.
\end{equation} 
This lower bound on the new physics scale is obviously far beyond what will be accessible at neutrino or collider experiments in the foreseeable future.  We note that this limit will be even stronger for larger values of $m_f$.
\end{subsection}

\begin{subsection}{${\cal O}_{\tilde{V}}$}
After electroweak symmetry breaking, $\phi$ acquires a vev and ${\cal O}_{\tilde{V}}$ gives a vertex
\begin{equation}
\frac{C_{\tilde{V}}}{\Lambda^2}{\cal O}_{\tilde{V}} \rightarrow \frac{-ig C_{\tilde{V}}v^2}{2\sqrt{2}\Lambda^2} \bar{\ell}_R \gamma^{\mu} f W_{\mu}^-
\end{equation}
which allows the decay $f\rightarrow e^-e^+\nu$ at tree level.  This process has  a width of
\begin{equation}
\Gamma(f\rightarrow e^+ e^- \nu)=\frac{|C_{\tilde{V}}|^2}{\Lambda^4} \frac{1}{1536 \pi^3} m_f^5. 
\end{equation}
\cite{Picciotto:2004rp} find that the lifetime for DM decaying to a final state that includes an $e^+e^-$ pair must satisfy 
\begin{equation}
\tau_{\tilde{V}} \simeq 5\times 10^{17} \mbox{yr} \frac{10 \mbox{ MeV}}{m_f}
\end{equation}
in order to not overproduce the 511 keV line observed by INTEGRAL \cite{Knodlseder:2005yq,Jean:2005af}.  From this constraint, we obtain the limits
\begin{eqnarray}
\frac{|C_{\tilde{V}}|^2}{\Lambda^4} &\lesssim& \frac{1}{(9.5\times 10^5\mbox{ TeV})^4} \mbox{ ($m_f=20$ MeV)}\nonumber\\
&\lesssim&\frac{1}{(2.4\times 10^6 \mbox{ TeV})^4} \mbox{ ($m_f=50$ MeV)}\\
&\lesssim&\frac{1}{(3.8\times 10^6\mbox{ TeV})^4} \mbox{ ($m_f=80$ MeV)},\nonumber
\end{eqnarray}
where the values of $m_f$ correspond to the values which we will eventually see are relevant for Super-K.  Again, we see that the constraints on the new physics scale for this operator are extremely strong, and, like the case for ${\cal O}_W$, this limit becomes stronger for larger values of $m_f$.

We will now use the very strong constraints which we have obtained on ${\cal O}_W$ and ${\cal O}_{\tilde{V}}$ to place constraints on our other operators via operator mixing.
\end{subsection}

\begin{subsection}{${\cal O}_{VR}$}
We now turn our attention to the operator ${\cal O}_{VR}$.  If $m_f>m_{\pi} + m_e$, ${\cal O}_{VR}$  can give the tree-level decay $f\rightarrow \pi^+ e^-$.  As we require $f$ to be very long-lived, this case is very strongly constrained.  Therefore, we only consider the mass range $m_f\lesssim m_{\pi}$.

However, ${\cal O}_{VR}$ can still induce a decay of the $f$ through the channel $f\rightarrow e^+e^-\nu$ via mixing into ${\cal O}_{\tilde{V}}$.  This mixing first occurs at one-loop level and is shown in Fig.~\ref{fig:oneloop}.  It must be noted that all of the fields in ${\cal O}_{VR}$ are right-handed.  However, the SM weak vertex in this diagram is entirely left-handed.  This implies that both the $u$ and $d$ quarks in the loop must flip chirality, and, thus, this diagram is suppressed by both the $u$ and $d$ quark Yukawa couplings.  This diagram is logarithmically divergent; we obtain, for the mixing of ${\cal O}_{VR}$ into ${\cal O}_{\tilde{V}}$
\begin{equation}
\frac{C_{\tilde{V}}(v)}{\Lambda^2}\sim\frac{C_{VR}(\Lambda)}{\Lambda^2}\frac{1}{(4\pi)^2}\frac{12 m_u m_d}{v^2}\ln{\frac{\Lambda^2}{m_f^2}}
\label{eq:ovr}
\end{equation}
which then results in the limits on the new physics scale
\begin{eqnarray}
\frac{|C_{VR}|^2}{\Lambda^4} &\lesssim& \frac{1}{(20\mbox{ TeV})^4} \mbox{ ($m_f=20$ MeV)}\nonumber\\
&\lesssim&\frac{1}{(50 \mbox{ TeV})^4} \mbox{ ($m_f=50$ MeV)}\label{eq:ovr15}
\\
&\lesssim&\frac{1}{(80\mbox{ TeV})^4} \mbox{ ($m_f=80$ MeV)}.\nonumber
\end{eqnarray}
\begin{figure}[h]
\centering
\includegraphics[width=80mm]{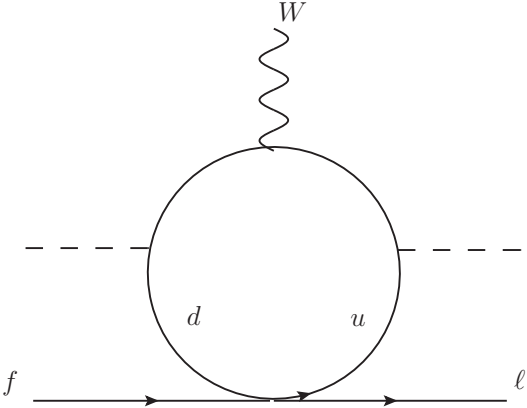}
\caption{One-loop mixing of ${\cal O}_{VR}$ into ${\cal O}_{\tilde{V}}$.  Here, the Higgs insertions which give the $u$ and $d$ quark chirality flips are explicitly shown.}
\label{fig:oneloop}
\end{figure}
However, this one-loop calculation does not accurately capture the contributions of loop momenta which are less than a few hundred MeV.  To estimate the effects of these low momenta, we also calculate the decay width for $f\rightarrow e^+e^-\nu$ where the decay goes through a virtual $\pi^+$.  This diagram receives suppression from both the electron and $f$ masses similar to the lepton mass dependence in the usual SM $\pi^+$ decay, as can be seen in the spin-averaged squared amplitude   
\begin{eqnarray}
\label{eq:ovr2}
&&\frac{1}{2}\displaystyle\sum_{spins}|{\cal M}|^2=\nonumber\\  &&\frac{|C_{VR}|^2}{\Lambda^4} \frac{G_F^2}{4} |V_{ud}|^2 f_{\pi}^4 m_e^2 m_f^2 \frac{q^2(m_f^2-q^2)}{(q^2-m_{\pi}^2)^2}
\end{eqnarray}
which gives the differential decay width
\begin{equation}
\frac{d\Gamma}{dq^2}=\frac{|C_{VR}|^2}{\Lambda^4}\frac{G_F^2|V_{ud}|^2f_{\pi}^4 m_e^2}{1024 \pi^3 m_f }\frac{q^2(m_f^2-q^2)^2}{(q^2-m_{\pi}^2)^2}.
\end{equation}
Integrating this expression over $q^2$, we obtain the limits
\begin{eqnarray}
\frac{|C_{VR}|^2}{\Lambda^4} &\lesssim& \frac{1}{(6\mbox{ TeV})^4} \mbox{ ($m_f=20$ MeV)}\nonumber\\
&\lesssim&\frac{1}{(20 \mbox{ TeV})^4} \mbox{ ($m_f=50$ MeV)}\\
&\lesssim&\frac{1}{(50\mbox{ TeV})^4} \mbox{ ($m_f=80$ MeV)}.\nonumber
\label{eq:ovr2.5}
\end{eqnarray}
\end{subsection}
As these limits are weaker than those obtained from the one-loop diagram, we take the one-loop results as our limits on the new physics scale for ${\cal O}_{VR}$.

As the limits in Eq. (\ref{eq:ovr15}) are substantially weaker than those which we have obtained for operators ${\cal O}_{W}$ and ${\cal O}_{\tilde{V}}$, we also consider other possible constraints on ${\cal O}_{VR}$.  First, we note that ${\cal O}_{VR}$ mixes into ${\cal O}_{W}$ at two-loop order; however, this mixing is suppressed by the $u$, $d$, and $e$ Yukawa couplings and is thus not competitive with the one-loop mixing into ${\cal O}_{\tilde{V}}$.  ${\cal O}_{VR}$ also induces a mass term coupling the $f$ to the SM $\nu$ at two-loop order, which can give the decays $f\rightarrow e^+e^-\nu$ and $f\rightarrow \nu\nu\bar{\nu}$; this mixing is similarly suppressed by three small Yukawa couplings.  Lastly, as $m_f\lesssim m_{\pi}$, one can consider the decay $\pi^+\rightarrow e^+ f$; searches for heavy neutrinos \cite{Britton:1992xv}, however, only constrain the new physics scale for ${\cal O}_{VR}$ to be greater than $O(10\mbox{ TeV})$.

\begin{subsection}{${\cal O}_{Sd}$, ${\cal O}_{Su}$, and ${\cal O}_{T}$}
Finally, we constrain ${\cal O}_{Sd}$, ${\cal O}_{Su}$, and ${\cal O}_{T}$ via their mixing into ${\cal O}_W$, which allows the decay $f\rightarrow \nu \gamma$.  ${\cal O}_{Sd}$ and ${\cal O}_{Su}$ mix into ${\cal O}_W$ at two-loop order.  However, as only one of the lepton fields in these operators is right-handed, these contributions are suppressed by only one power of a small Yukawa coupling.  We obtain an order-of-magnitude estimate of the mixing of these operators into ${\cal O}_W$,
\begin{eqnarray}
\frac{C_W(v)}{\Lambda^2}&\sim & \frac{C_{Su,Sd}(\Lambda)}{\Lambda^2} \frac{1}{(4 \pi)^4}  \frac{g^2 m_{u,d}}{v} \ln{\left(\frac{\Lambda^2}{v^2}\right)}\\ &\sim &\frac{C_{Su,Sd}(\Lambda)}{\Lambda^2}\times 10^{-9}.
\label{eq:osu}
\end{eqnarray}
This suppression is not sufficient to make $f$ long lived.  Again assuming that this decay does not happen at a rate faster than $O((10^{26}\mbox{ s})^{-1})$, we obtain the order-of-magnitude limit
\begin{equation}
\frac{C_{Su,Sd}(\Lambda)}{\Lambda^2} < O\left(\frac{1}{(10^3 \mbox{ TeV})^2}\right).
\end{equation}

${\cal O}_{T}$, on the other hand, mixes into ${\cal O}_W$ at one-loop order, again with only one suppression by a small Yukawa coupling.  Thus, it will be even more strongly constrained than ${\cal O}_{Sd}$ and ${\cal O}_{Su}$.

For the rest of this work, we consider only our most weakly constrained operator, ${\cal O}_{VR}$.
\end{subsection}

\section{Signals in Neutrino Experiments}
\label{sec:nus}
We now consider the observability of ${\cal O}_{VR}$ in neutrino experiments.  In order to estimate the reach of a neutrino detector to observe the process $\bar{f} p\rightarrow e^+ n$, it is first useful to estimate the possible flux of $f$ at the Earth's surface.  The DM mass density in our neighborhood is thought to be roughly $0.3 \mbox{ GeV}/\mbox{cm}^3$ \cite{Caldwell:1981rj}, and its velocity relative the the Earth $v_f$ is approximately $O(10^{-3})$ (for $c=1$)\cite{Kamionkowski:1997xg}.  If we assume that $\bar{f}$ comprises all DM, we can thus estimate its flux
\begin{equation}
\Phi_{\bar{f}}\sim\frac{0.3\mbox{ GeV}/\mbox{cm}^3}{m_f} v_f c \sim (10^{10},10^9,10^8) /\mbox{cm}^2\mbox{s}
\label{eq:flux}
\end{equation}
for $m_f=(1, 10, 100)$ MeV, respectively.  We can now compare this with the limit on the relic supernova $\bar{\nu}_e$ flux obtained by Super-K \cite{Malek:2002ns} of $<1.2\bar{\nu}_e/\mbox {cm}^2 \mbox{s}$.  They obtain this number by fitting the overall normalizations of their background and signal energy distributions to data, as their background and signal shapes are sufficiently distinct.  As the $e^+$ in $\bar{f} p\rightarrow e^+ n$ is essentially monoenergetic, we assume that this process could be distinguished from background at least as well.  (Because their signal shape is different from that for $\bar{f} p\rightarrow e^+ n$, it is possible that our results are slightly overly optimistic for certain ranges of $m_f$.  However, it is unlikely that this will affect the lower bounds on the new physics scale by more than a few tens of percent.)  As their flux limit is $8$ to $10$ orders of magnitude smaller than that in Eq. (\ref{eq:flux}), we can constrain the cross-section $\sigma_{{\cal O}}$ for our process 
\begin{equation}
\sigma_{{\cal O}}\simeq \frac{1}{16\pi |v_f|} \frac{ |C_{VR}|^2}{\Lambda^4}  m_f^2 (|f_1|^2+3|g_1|^2)
\end{equation}
to be $8$ to $10$ orders of magnitude smaller than the SM neutrino cross-section $\sigma_{SM}$
\begin{equation}
\sigma_{SM} \simeq \frac{G_F^2}{\pi} E_{\nu}^2 (|f_1|^2+3|g_1|^2).
\end{equation}
Here $f_1\simeq 1$ and $g_1\simeq -1.27$ are the nucelon form factors \cite{Strumia:2003zx}.
We thus obtain for the ratio of these two cross-sections
\begin{equation}
\frac{|C_{VR}|^2 v^4}{8 |v_f| \Lambda^4}\leq \frac{1.2/\mbox{cm}^2\mbox{s}}{(0.3\mbox{ GeV}/\mbox{cm}^3) |v_f| c/m_f},
\end{equation}
from which we obtain
\begin{eqnarray}
\label{eq:results}
\frac{|C_{VR}|^2}{\Lambda^4} &\lesssim& \frac{1}{(120\mbox{ TeV})^4} \mbox{ ($m_f=20$ MeV)}\nonumber\\
&\lesssim&\frac{1}{(90 \mbox{ TeV})^4} \mbox{ ($m_f=50$ MeV)}\\
&\lesssim&\frac{1}{(80\mbox{ TeV})^4} \mbox{ ($m_f=80$ MeV)}.\nonumber
\end{eqnarray}
for approximate lower bounds on the scale of new physics for ${\cal O}_{VR}$.  We note that the values are tighter for smaller $m_f$ because the assumed $\bar{f}$ flux is inversely proportional to $m_f$.  Of course, scenarios in which $\bar{f}$ comprises only some small fraction of DM would be more weakly constrained.
\section{Discussion}
\label{sec:dis}
Here, we briefly discuss a few characteristics of the interaction ${\cal O}_{VR}$ and our DM candidate $f$.  One question not yet addressed here is that of the $f$ (or $\bar{f}$) relic density.  The new physics scale relevant for ${\cal O}_{VR}$, $\gtrsim O(100\mbox{ TeV})$, is much too high to give $f$ a relic density compatible with observation.  Thus, some other interaction, such as one which allows the $f$ to annihilate to $e^+e^-$ pairs or neutrinos, must be postulated to exist in addition to ${\cal O}_{VR}$.  Such interactions must have physics scales that are very low, on the order of a few GeV.

Next, we very briefly mention the possible applicability of ${\cal O}_{VR}$ to particular models.  One may note that ${\cal O}_{VR}$ appears very similar to a right-handed neutrino interaction.  In fact, $f$, being an SM singlet, has the same quantum numbers as a right-handed neutrino.  One can write down a dimension-four mass operator, $\bar{L}\tilde{\phi}f$, which allows the $f$ to mix with the SM neutrino and gives the decays $f\rightarrow \nu e^+ e^-$ and $f\rightarrow \nu\nu\bar{\nu}$.  As we want the $f$ to be long-lived, we wish to postulate some symmetry which will disallow this operator.  Although it is possible to impose a symmetry which excludes this four-dimensional operator but still allows ${\cal O}_{VR}$, doing so requires introducing at least an additional Higgs doublet to the SM.  A full exploration of model-building with ${\cal O}_{VR}$ is beyond the scope of this work, but, given the form of the interaction, an investigation of left-right-symmetric models may be a worthwhile endeavor.

Lastly, we have not specified whether $f$ is a Dirac or Majorana fermion.  We briefly note that if $f$ is Majorana, it necessarily violates lepton number.  

\section{Conclusions}
\label{sec:con}
Here, we have investigated the possibility of the direct detection of DM in neutrino experiments via a model-independent analysis.  We have considered operators which contribute to the interaction $\bar{f} p\rightarrow e^+ n$ and placed limits on the coefficients of these operators using DM lifetime and decays.  There exists one operator which is comparatively weakly constrained for the case where $m_f\lesssim O(100\mbox{ MeV})$.  We find that Super-K can probe the scale of new physics for this operator up to $O(100\mbox{ TeV})$.

We draw two main conclusions from this work.  The first is that, given our lack of knowledge of DM interactions with SM particles, ``nontraditional'' possibilities should be considered.  The inelasticity of the interaction $\bar{f} p\rightarrow e^+ n$ allows one to probe the region $m_f\sim 100$ MeV, a range not usually accessible to DM direct-detection experiments.  Second, we find that the scale which can be probed for such DM is very impressive, $O(100\mbox{ TeV})$, far beyond the scales usually accessible in collider experiments.  

Given these results, it may be fruitful to consider how this analysis can be expanded to other inelastic interactions, such as $f_1 N \rightarrow f_2  N$, where both $f_1$ and $f_2$ are invisible particles and where $f_2$ (which could be either a neutrino or a BSM particle) is lighter than $f_1$.  In this case, this interaction could conceivably produce distinctive signatures in either neutrino experiments or traditional DM direct-detection experiments; an analysis of such an interaction could possibly be relevant for models of Inelastic DM \cite{TuckerSmith:2001hy} or Exciting DM \cite{Finkbeiner:2007kk}.  The investigation of these interactions with an invisible particle in the final state is left for future study.

\bigskip 
\begin{acknowledgments}
First, I would like to thank the organizers of DPF 2009 for arranging such an excellent conference.  I also owe many thanks to A. Soni, my collaborator on the project on which this talk was largely based.  I would also like to thank M. Wise for his calculation of the $f\rightarrow e^+e^-\nu$ decay width as well as many additional suggestions which greatly improved the project.  Lastly, I would like to thank H. Davoudiasl, S. Dawson, S. Gopalakrishna, W. Marciano, C. Sturm, and M. Ramsey-Musolf for their helpful advice and discussions on many topics related to this analysis.  This work is supported under US DOE contract No. DE-AC02-98CH10886.
\end{acknowledgments}

\bigskip 

\end{document}